\documentclass[review]{elsarticle}

\biboptions{sort&compress}

\journal{Nucl. Instr. and Meth. A}

\usepackage[dvipdfmx]{hyperref}

\newcommand{\gsim}{\hbox{ \raise3pt\hbox to 0pt{$>$}\raise-3pt\hbox{$\sim$} }}
\newcommand{\lsim}{\hbox{ \raise3pt\hbox to 0pt{$<$}\raise-3pt\hbox{$\sim$} }}
\newcommand{\del}{\ifmmode{\nabla}         \else{$\nabla$ }               \fi}

\newcommand{\figdir}{.}

\usepackage{graphicx}

\usepackage{comment}

\begin{document}

\begin{frontmatter}

\title{A novel technique for the measurement of the avalanche
 fluctuation of gaseous detectors}

%
%
\author[3]{M.~Kobayashi\corref{cor1}}
     \ead{makoto.kobayashi.exp@kek.jp}
     \cortext[cor1]{Corresponding author.
                           Tel.: +81 29 864 5376; fax: +81 29 864 2580.}
\author[7]{T.~Ogawa}
\author[8]{T.~Kawaguchi}
\author[3]{K.~Fujii}
\author[1]{T.~Fusayasu}
\author[1]{K.~Ikematsu}
\author[4]{Y.~Kato}
\author[10]{S.~Kawada}
\author[3]{T.~Matsuda}
\author[5]{R.D.~Settles}
\author[1]{A.~Sugiyama}
\author[8]{T.~Takahashi}
\author[3]{J.~Tian}
\author[6]{T.~Watanabe}
\author[9]{R.~Yonamine}
%
%
\address[3]{High Energy Accelerator Research Organization (KEK), Tsukuba 305-0801, Japan}
\address[7]{The Graduate University for Advanced Studies (Sokendai), Tsukuba 305-0801, Japan}
\address[8]{Hiroshima University, Higashi-Hiroshima 739-8530, Japan}
\address[1]{Saga University, Saga 840-8502, Japan}
\address[4]{Kinki University, Higashi-Osaka 577-8502, Japan}
\address[10]{Deutsches Elektronen-Synchrotron (DESY), D-22607 Hamburg, Germany}
\address[5]{Max Planck Institute for Physics, DE-80805 Munich, Germany}
\address[6]{Kogakuin University, Shinjuku 163-8677, Japan}
\address[9]{Universit$\acute{e}$ Libre de Bruxelles, 1050 Brussels, Belgium}

%
%
%
%
\begin{abstract}
We have developed a novel technique for the measurement of the avalanche
fluctuation of gaseous detectors using a UV laser.
The technique is simple and requires a short data-taking time of about
ten minutes. Furthermore, it is applicable for relatively low gas gains.
Our experimental setup as well as the measurement principle,
and the results obtained with a stack of Gas Electron Multipliers (GEMs)
operated in several gas mixtures are presented.
\end{abstract}

%
%
%
%

\end{frontmatter}


\section{Introduction}

Gas amplification of the electrons created by X-rays, UV photons, or
charged particles plays an essential role in their detection with
gaseous detectors.
It acts as  ``preamplifier'' with a sufficient gain. However, its gain
fluctuates because of
avalanche statistics, thereby degrading the energy resolution for
monochromatic X-rays.
For large Time Projection Chambers (TPCs) the azimuthal spatial
resolution at long drift distances is partly limited by the relative variance
of the gas gain for single drift electrons~\cite{Kobayashi1,Kobayashi2}.
Historically many analytical models have been proposed
to understand the fluctuation
(see, for example,  Ref.~\cite{Alkhazov} and reference cited therein).
A Monte-Carlo simulation approach is also available these days~\cite{Thomas}.

Conventionally, avalanche fluctuations are estimated from the
gas-amplified charge spectrum
for single electrons created by a UV lamp or a laser.
This method is, however, not easy because of electronic noise interference,
especially for low gas gains.
We have developed a novel technique for the measurement of the relative
variance of avalanche fluctuation ($f$) using laser-induced tracks,
exploiting the fixed cluster size of unity for each ionization act along the tracks.
The primary electrons are multiplied by a gas amplification device,
and then collected by readout pad rows arranged along the laser beam.
The signal charges on adjacent pad rows are compared for each laser shot.
The value of $f$ is estimated from the width of the distribution of their differences
using a straightforward relation. The technique is relatively simple and
the typical data-taking time is ten minutes.

\section{Experimental setup and measurement principle}

The schematic view of the experimental setup is shown in Fig.~1.
A pulsed UV laser beam is injected into a chamber box, located at the
laser focal point, through a quartz window.
The chamber box contains a pair of GEM foils~\cite{Sauli}, a readout plane right
behind the GEMs, and a mesh drift electrode.
The readout plane is paved over with $\sim$ 5.3 mm wide
pad rows arranged normally to the laser beam.
The GEM foil is a laser-etched 100-$\mu$m thick liquid crystal polymer film
with copper electrodes.
The hole diameter (pitch) is 70 $\mu$m (140 $\mu$m).
The electrons created along a laser-induced track are amplified by GEMs,
and then collected by the pad rows, after 1.6 cm of drift.
The signal charge on each pad is digitized by a flash ADC 
for every laser shot.
An $^{55}$Fe source is placed under the chamber box for calibration (see below). 

The relative variance of gas gain ($f$) is estimated from the difference of
the signal charges
$Q_1$ and $Q_2$ on adjacent pad rows (see Fig.~2) for each laser shot, using the relation:
\begin{eqnarray}
1 + f &=& \frac{\left< n \right>}{2} \cdot \left< R^2 \right>\nonumber \\
 &\equiv& \frac{\left< n \right>}{2} \cdot
 \left< \left( \frac{Q_1 - Q_2}{\left< Q \right>} \right)^2
 \right> 
\end{eqnarray}
with $\left< n \right>$ being the average number of primary electrons per pad row created by
the laser beam,
and $\left< Q \right>$, the normalized average charge on a pad row,
for a series of laser shots.
See Appendix for the derivation of Eq.~(1).
The value of $\left< n \right>$ is calculated from the average signal
charge obtained with the laser
and that given by the $^{55}$Fe source (see Fig.~3)\footnote{
The Penning effect is not taken into account in the estimation of
the average number of electrons created by 5.9 keV X-rays from $^{55}$Fe.
},
and is typically about 250.
A Poisson distribution is assumed for $n$ with its average common to
the nearby pad rows for the {\em same} single laser shot.
This technique using the signal charges on a couple of adjacent pad
rows is tolerant of the variation of the laser intensity (see Appendix),
and insensitive to the dominant common-mode electronic noise on the pad rows (see Fig.~4).

\section{Results}

Figs.~5-7 show the distributions of $\sqrt{\left< n \right> / 2} \cdot R$
obtained with three gas mixtures: Ar-methane (5\%), 
Ar-CF$_4$ (3\%)-isobutane (2\%) and Ne-isobutane (5\%).
The first two gas mixtures are candidates for the TPC of the linear collider experiment,
while the third mixture is expected to provide smaller gas gain fluctuations. 
The gas pressure was kept around 1 atmosphere.
We used combinations of every other pad rows, instead of neighboring ones
in order to avoid the possible correlation caused by the charge sharing
due to the diffusion of amplified electrons in the transfer/induction gaps,
and electronic crosstalk in the signal cables to the readout electronics.
The distribution is almost normal, and its variance obtained by
Gaussian fitting gives the estimator for the value of $1 + f$.  
These results are preliminary and the errors assigned are statistical only.

The values of $f$ measured with the argon-based gases are close to the indirect estimation from
the resolution degradation with increasing drift distance observed with a 
prototype TPC for the linear collider experiment~\cite{Kobayashi1,Kobayashi2}.
The measurement with the neon-based mixture is consistent with that obtained with
Micromegas using the charge spectrum for single electrons~\cite{Thomas}.

\section{Conclusion}

Our technique using laser-induced tracks for the measurement of
avalanche fluctuation is relatively simple and
easier than observing the charge spectrum for single electrons,
and requires no assumption for the functional form of the charge spectrum.
The typical data-taking time is about 10 minutes, corresponding to 10 000
laser shots.
The resultant values of the relative variance ($f$) are consistent with
other measurements.
It is worth mentioning that this technique relies on the absence of
secondary ionizations,
i.e. cluster size = 1, in the 2-photon ionization process by the laser photons.
It is, therefore, not applicable for charged particle tracks because of
large Landau fluctuations.

The present experiment was carried out  using a readout module of a
prototype TPC for the linear collider experiment~\cite{Yonamine},
for the proof of principle of our technique.
Actually, many pad rows, each consisting of small pads ($\sim$ 1.2 mm
$\times$ 5.3 mm)
equipped with individual readout, are not necessary for the measurement. 
We are now preparing a smaller dedicated chamber with several readout pads or strips
in a moderate size,
in order to measure the gas gain fluctuations for different gas amplification devices
such as Micromegas~\cite{Giomataris} and counting gases. 

\begin{figure}[htbp]
\begin{center}
\hspace{10mm}
\includegraphics*[scale=0.5]{\figdir/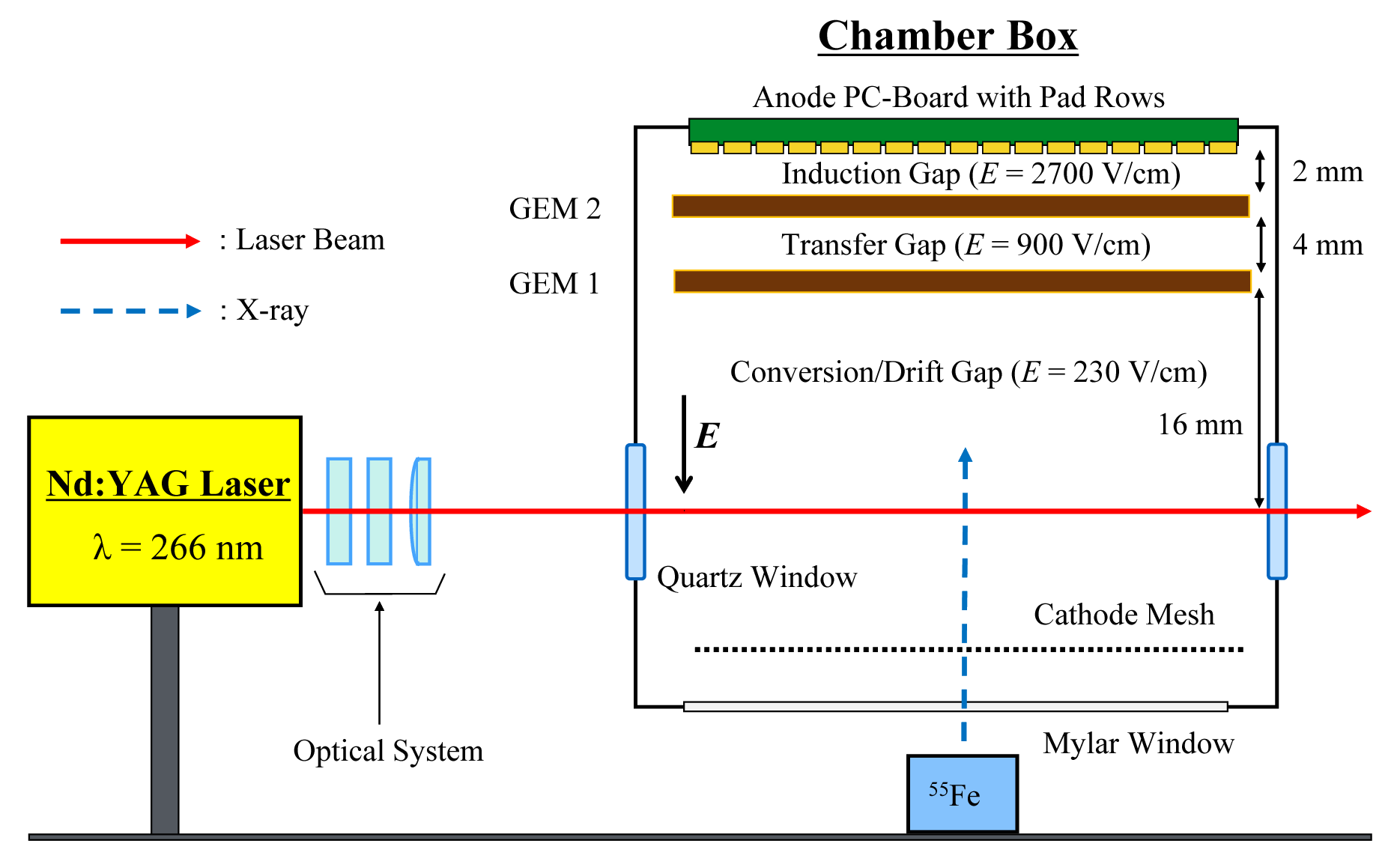}
\end{center}
\vspace{-5mm}
\caption{\label{fig1}
\footnotesize Schematic view of the experimental setup.  
}
\end{figure}
\begin{figure}[htbp]
\begin{center}
\hspace{10mm}
\includegraphics*[scale=0.7]{\figdir/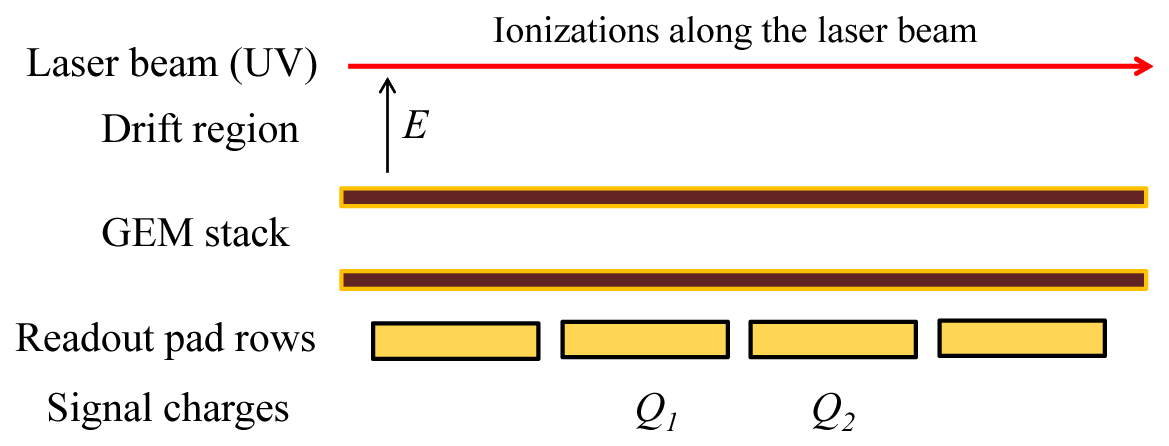}
\end{center}
\vspace{-5mm}
\caption{\label{fig2}
\footnotesize Definition of $Q_1$ and $Q_2$.  
}
\end{figure}
\begin{figure}[htbp]
\begin{center}
\hspace{10mm}
\includegraphics*[scale=0.55]{\figdir/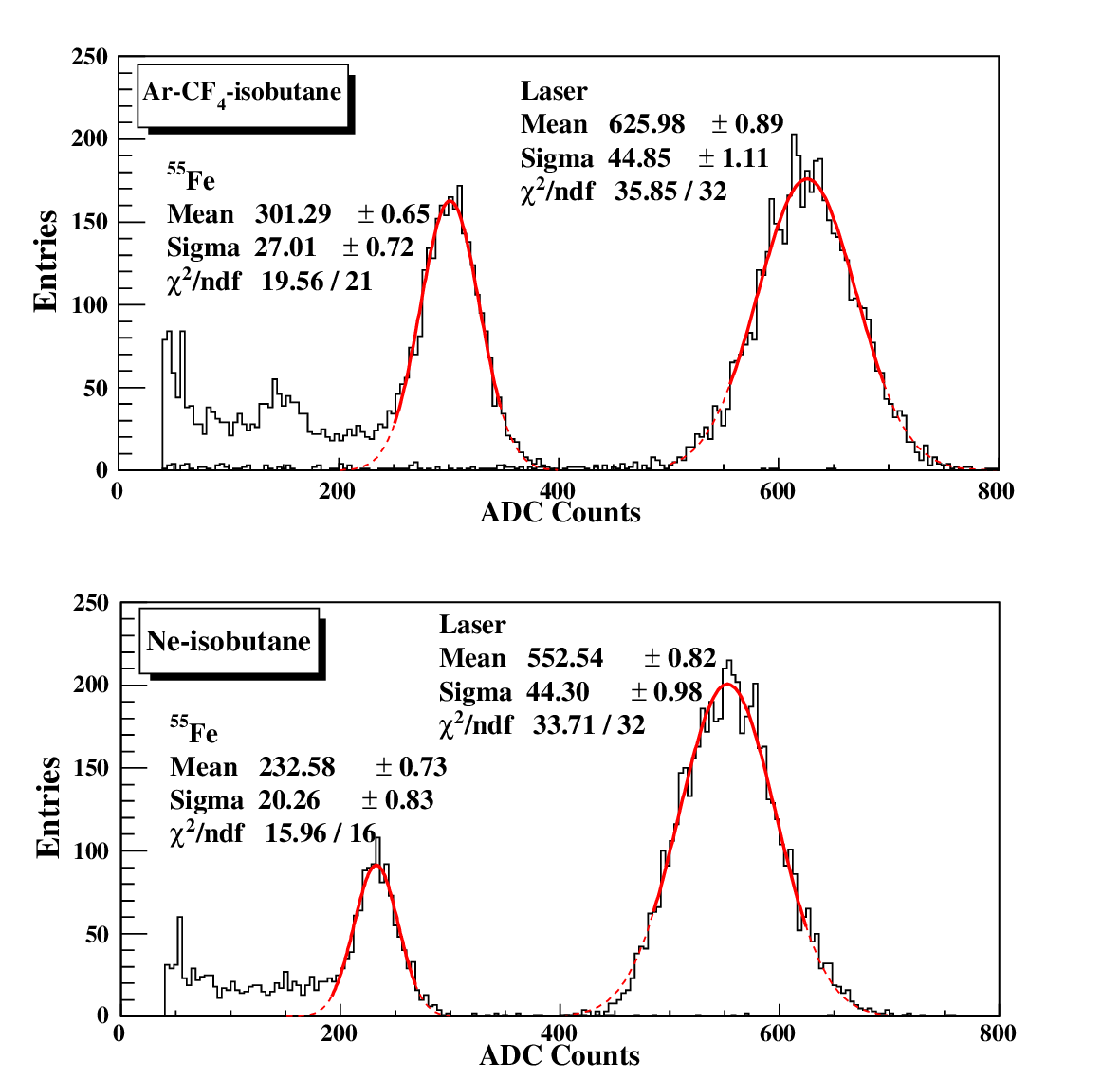}
\end{center}
\vspace{-5mm}
\caption{\label{fig3}
\footnotesize Signal charge distributions for the $^{55}$Fe source and
 the laser beam measured with a couple of pad rows: the upper panel for
an argon-based gas mixture (Ar-CF$_4$ (3\%)-isobutane (2\%))
and the lower for a neon-based mixture (Ne-isobutane (5\%)).  
}
\end{figure}
\begin{figure}[htbp]
\begin{center}
\hspace{10mm}
\includegraphics*[scale=0.4]{\figdir/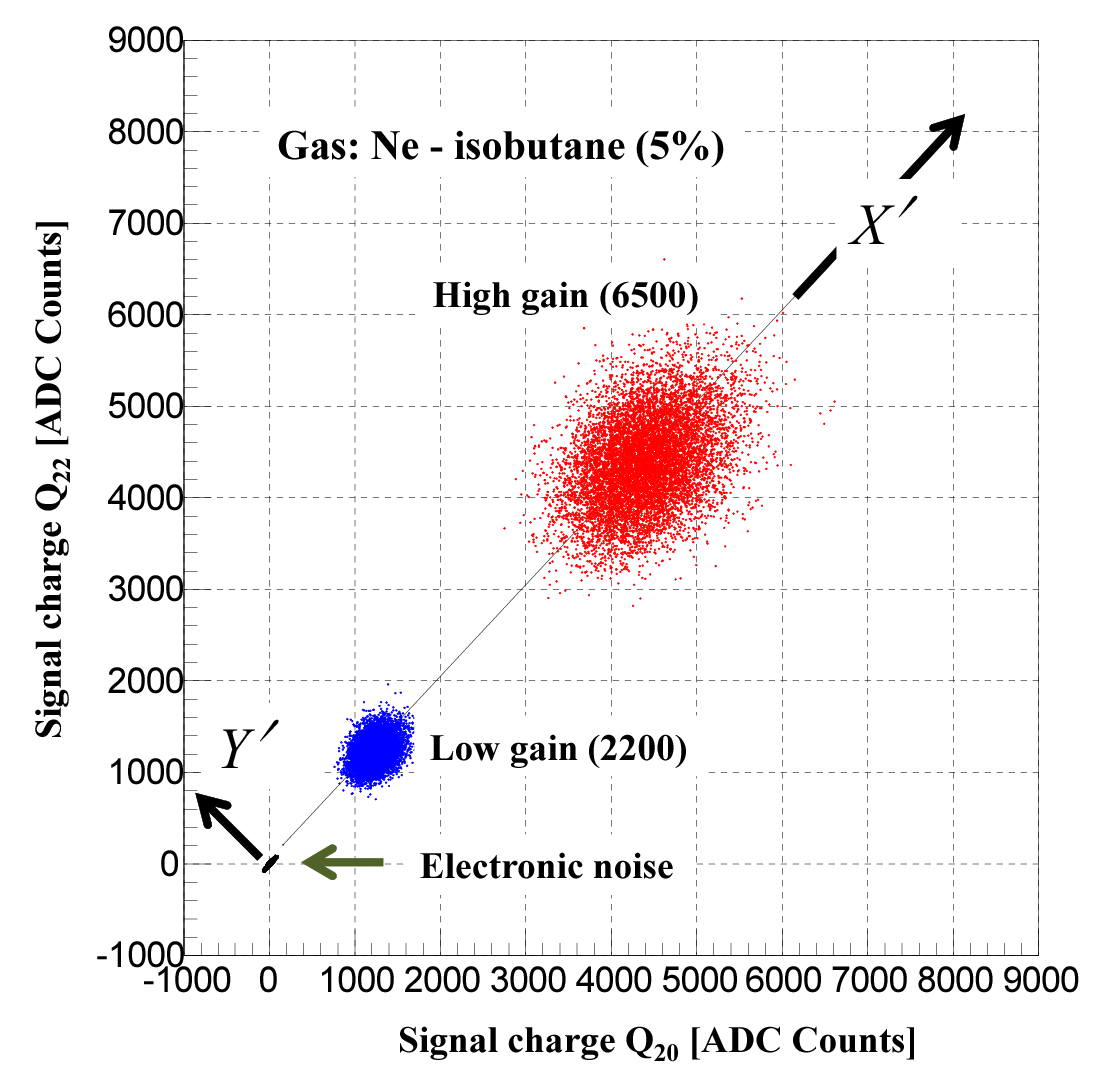}
\end{center}
\vspace{-5mm}
\caption{\label{fig4}
\footnotesize Example of the correlation of the laser signal charges
measured by two pad rows (pad row \#20 and \#22).
They are correlated because of the variation of the laser
intensity: drift and/or shot-to-shot fluctuations.
It is therefore important to make use of the projection on the
$Y^{\prime}$-axis, which is proportional to $Q_{22}-Q_{20}$,
in order to avoid the influence of the laser intensity variation (see Appendix). 
The electronic noise charges are strongly correlated, indicating 
that the common mode noise is dominant. 
}
\end{figure}
\begin{figure}[htbp]
\begin{center}
\hspace{10mm}
\includegraphics*[scale=0.7]{\figdir/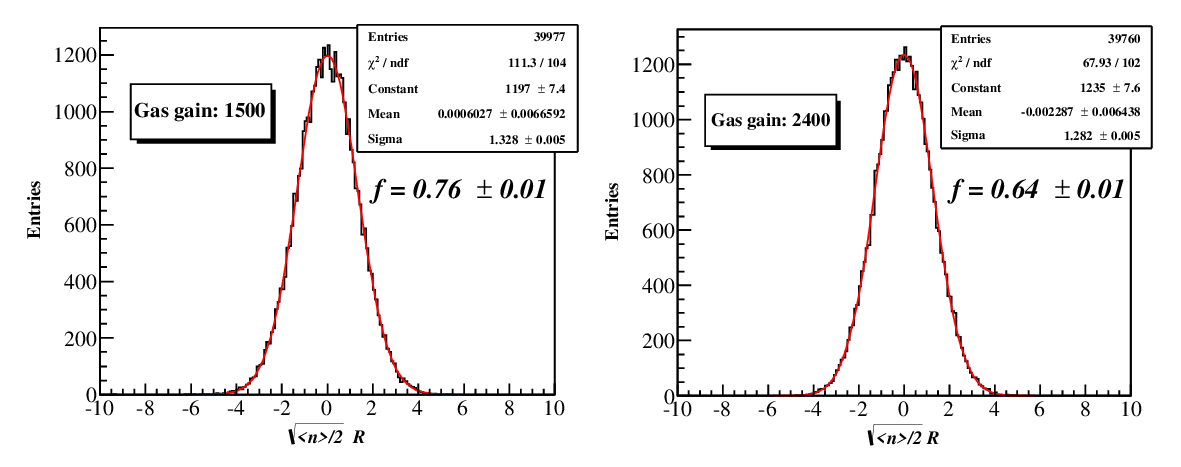}
\end{center}
\vspace{-5mm}
\caption{\label{fig5}
\footnotesize Distributions of $\sqrt{\left< n \right> / 2} \cdot R$
for an Ar-methane (5\%) mixture for a lower gas gain (left) and a higher gas
 gain (right).
}
\end{figure}
\begin{figure}[htbp]
\begin{center}
\hspace{10mm}
\includegraphics*[scale=0.7]{\figdir/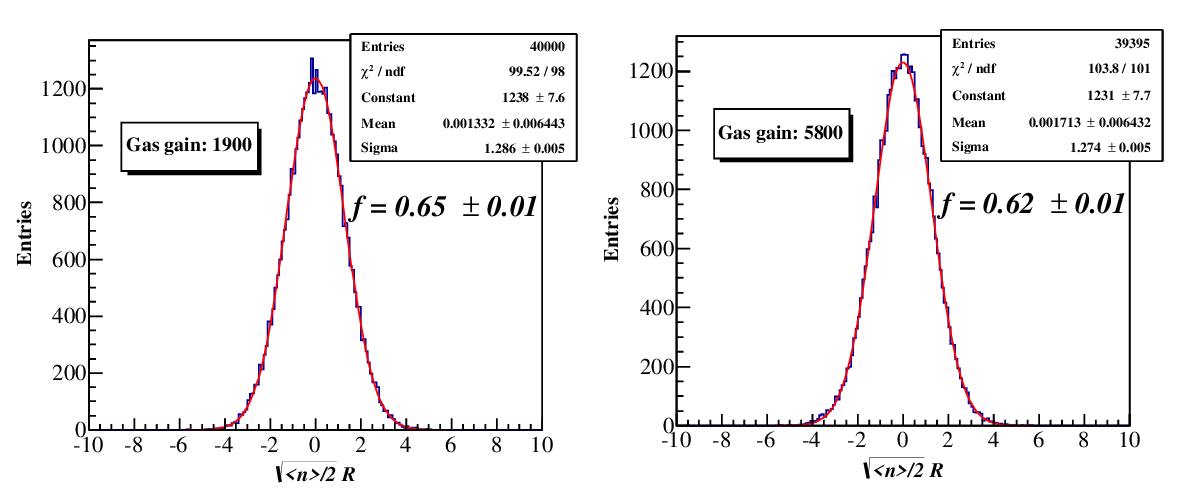}
\end{center}
\vspace{-5mm}
\caption{\label{fig6}
\footnotesize Distributions of $\sqrt{\left< n \right> / 2} \cdot R$
for an Ar-CF$_4$ (3\%)-isobutane (2\%) mixture for a lower gas gain (left) and a higher gas
 gain (right).
}
\end{figure}
\begin{figure}[htbp]
\begin{center}
\hspace{10mm}
\includegraphics*[scale=0.7]{\figdir/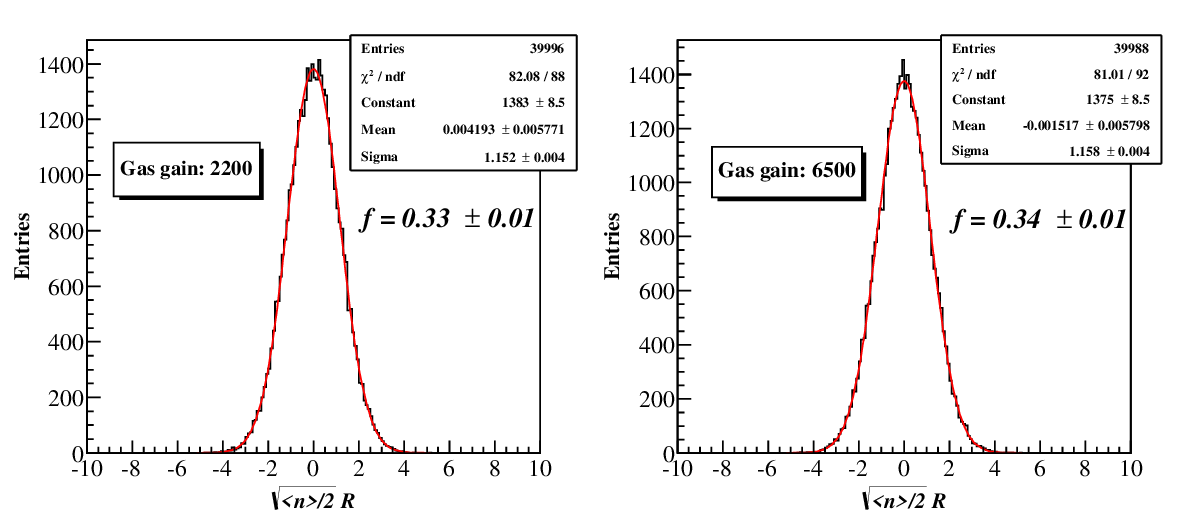}
\end{center}
\vspace{-5mm}
\caption{\label{fig7}
\footnotesize Distributions of $\sqrt{\left< n \right> / 2} \cdot R$
for a Ne-isobutane (5\%) mixture for a lower gas gain (left) and a higher gas
 gain (right).
}
\end{figure}

\newpage

\section*{\nonumber Acknowledgments}
This work was partly supported by Grant-in-Aid for Specially Promoted Research
Grant no. 23000002 of Japan Society for the Promotion of Science.

\appendix
\section{Derivation of Eq.~(1)}

Let us define $n$ to be the total number of primary electrons per pad row created by the laser
beam, $g_{\rm i}$ to be the gas gain for the $i$-th liberated electron,
and $N$ be the total number of amplified electrons collected by a pad row.
Then
\begin{eqnarray}
N = \sum_{i=1}^{n} \; g_{\rm i} &=& g_1 + g_2 + \;, \cdot \cdot \cdot \;,+ g_{\rm n} \\
\left< N \right> &=& \left< n \right> \cdot \left< g \right> \\
\sigma_{\rm N}^2 &\equiv& \left< \left( N - \left< N \right> \right)^2 \right> \\
&=& \left< \Bigl( \bigl( (g_1 - \left< g \right>) + (g_2 - \left< g\right>) 
\; + \;, \cdot \cdot \cdot \;, + \; (g_{\rm n} - \left< g \right>) \bigr)   
   + n \cdot \left< g \right> - \left< n \right> \cdot \left< g \right>
	    \Bigr)^2 \right> \\
&=& \left< n \cdot \left( g - \left< g \right> \right)^2 \right>
 + \left< g \right>^2 \cdot \left< \left( n - \left< n \right> \right)^2
 \right> \\
&=& \left< n \right> \cdot \sigma_{\rm g}^2 + \left< g \right>^2 \cdot
 \sigma_{\rm n}^2 \\
&=& \left< n \right> \cdot \left< g \right>^2 \cdot (1 + f)
\end{eqnarray}
with $f$ being the relative variance of the gas gain for single
drift electrons ($\equiv \sigma_{\rm g}^2 / \left< g \right>^2$).
In Eq.~(A.7) $\sigma_{\rm n}^2$ is replaced by $\left < n \right >$,
assuming Poisson statistics for the number of initial ionizations.

In practice, $\left< n \right>$ varies with time because of the
laser intensity variation due to drift and/or shot-to-shot fluctuations,
and the signal charges,  proportional to $n$, on different pad rows
are correlated (see Fig. 4):
\begin{equation}
\left< n \right> \equiv \left<\left< n \right>\right> + \delta \left<n\right>
\end{equation}
and
\begin{equation} 
n \equiv \left<\left<n\right>\right> + \delta \left<n\right> + \delta n
\end{equation}
where $\left< n \right>$ is the average number of electrons
per pad row created by a single laser shot
and $\left<\left< n \right>\right>$ , the average of 
$\left< n \right>$ during a data taking period.
Therefore


\begin{eqnarray}
\left< \sigma_{\rm N}^2\right> &\equiv& \left< \bigl( N - 
                    \left< \left< N \right> \right> \bigr)^2 \right> \\
&=& \left<  \bigl( N - \left< N \right> + \left< N \right> -
     \left< \left< N \right> \right> \bigr)^2 \right> \\
&=&  \left< \bigl( N - \left< N \right> \bigr)^2 \right>
      +  \left< \bigl( \left< N \right> - \left< \left< N \right> \right> \bigr)^2 \right> \\
&=& \left< \left< n \right> \right> \cdot \left< g \right>^2 \cdot (1+f)
 + \left< g \right>^2 \cdot \left< \bigl( \left< n \right> -
			      \left< \left< n \right> \right> \bigr)^2 \right> \\
&=& \left< \left< n \right> \right> \cdot \left< g \right>^2 \cdot (1+f)
     + \left< g \right> ^2 \cdot \left< \left( \delta
     \left< n \right> \right) ^2 \right> 
\end{eqnarray}
where $\left< N \right>$ is the average number of gas-amplified
electrons per pad row for a single laser shot
and $\left<\left< N \right>\right>$ , the average of 
$\left< N \right>$ during the data taking period.
Obviously
\begin{eqnarray}
\left< N \right> &=& \left< g \right> \cdot \left< n \right> \\
\left<\left< N \right>\right> &=& \left< g \right> \cdot \left<\left< n \right>\right>\;. 
\end{eqnarray}

The variance of the difference in the amplified number of electrons
detected by two nearby pad rows, averaged over the data taking period,
is given by
\begin{eqnarray}
\left< \left( N_1 - N_2 \right)^2 \right> 
&=& \left< \Bigl(\bigl( N_1 - \left< \left< N \right> \right> \bigr) - 
   \bigl( N_2 - \left< \left< N \right> \right> \bigr) \Bigr)^2 \right> \\ 
&=& \left<\bigl(N_1-\left<\left<N\right>\right>\bigr)^2 \right>
      + \left<\bigl(N_2-\left<\left<N\right>\right>\bigr)^2 \right>
      -2 \cdot \Bigl< \bigl( N_1 - \left<\left<N\right>\right>\bigr)
         \cdot \bigl( N_2 - \left<\left<N\right>\right>\bigr)\Bigr> \\   
&=& 2 \cdot \left< \left< n \right> \right> \cdot \left< g \right>^2
 \cdot (1+f)
 + 2 \cdot \left< g \right>^2 \cdot \left< \left( \delta\left< n \right>
					   \right)^2 \right>
 - 2 \cdot \Bigl< \bigl(N_1-\left<\left< N \right>\right> \bigr)
     \cdot \bigl(N_2-\left<\left< N \right>\right> \bigr) \Bigr> \\ 
&=& 2 \cdot \left< \left< n \right> \right> \cdot \left< g \right>^2 \cdot (1+f)
 + 2 \cdot \left< g \right>^2 \cdot \left< \left( \delta\left< n \right>
					   \right)^2 \right> \nonumber \\ 
&&\hspace{10em} - 2 \cdot \Bigl< \bigl(N_1-\left<N\right> + \left<N\right>-\left<\left< N \right>\right> \bigr)
     \cdot \bigl(N_2-\left<N\right>+\left<N\right> - \left<\left< N \right>\right> \bigr) \Bigr> \\
&=& 2 \cdot \left< \left< n \right> \right> \cdot \left< g \right>^2
 \cdot (1+f)
 + 2 \cdot \left< g \right>^2 \cdot \left< \left( \delta\left< n \right>
					   \right)^2 \right>
 - 2 \cdot \left< \bigl( \left< N \right> - \left< \left< N \right>\right> \bigr)^2 \right> \\
&=& 2 \cdot \left< \left< n \right> \right> \cdot \left< g \right>^2
 \cdot (1+f)
 + 2 \cdot \left< g \right>^2 \cdot \left< \left( \delta\left< n \right>
					   \right)^2 \right> 
 - 2 \cdot \left< g \right> ^2 \cdot \left< \bigl( \left< n \right>
                                - \left< \left< n \right>\right> \bigr)^2 \right> \\
&=&  2 \cdot \left< \left< n \right> \right> \cdot \left< g \right>^2
 \cdot (1+f)
\end{eqnarray}
assuming $\left<n_1\right> = \left< n_2 \right> = \left< n \right>$ and
$\left<N_1\right> = \left< N_2 \right> = \left< N \right>$
for the adjacent pad rows.
The influence of the laser intensity variation
($\left< \left( \delta \left< n \right> \right) ^2 \right>$)
can thus be eliminated
by measuring $N_1 - N_2$ on an event-by-event (shot-by-shot) basis.

Accordingly
\begin{eqnarray}
1 + f &=& \frac{1}{2} \cdot \frac{\left<\left( N_1 - N_2 \right)^2\right>}
{\left< g \right>^2 \cdot \left<\left< n \right> \right>} \\
&=& \frac{\left<\left< n \right>\right>}{2} \cdot 
\frac{\left< \left( N_1 - N_2 \right)^2
\right>}{\left<\left< N \right>\right>^2} \\
&=& \frac{\left<\left< n \right>\right>}{2} \cdot 
\frac{\left< \left( Q_1 - Q_2 \right)^2
\right>}{\left<\left< Q \right>\right>^2}
\end{eqnarray}
where $Q$ represents the charge, proportional to $N$, recorded by the
readout electronics.

In the calculation the diffusion of amplified electrons in the
transfer/induction gaps is neglected.
Although the diffusion is small compared to the pad-row width
it causes the charge sharing between neighboring pad rows
and reduces the fluctuation of the signal charge on a pad row.
Therefore Eqs.~(A.7) and (A.23) slightly overestimate
the variances.


\newpage

\begin{frontmatter}



\title{
  Addendum to ``A novel technique for the measurement of the avalanche
 fluctuation of gaseous detectors''
  [Nucl. Instrum. Methods Phys. Res. A 845 (2017) 236-240]
}
%
%
%
%
%
\author[3]{M.~Kobayashi\corref{cor1}}
     \ead{makoto.kobayashi.exp@kek.jp}
     \cortext[cor1]{Corresponding author.
                           Tel.: +81 29 864 5376; fax: +81 29 864 2580.}
\author[7]{T.~Ogawa}
\author[8]{T.~Kawaguchi}
\author[3]{K.~Fujii}
\author[1]{T.~Fusayasu}
\author[1]{K.~Ikematsu}
\author[4]{Y.~Kato}
\author[10]{S.~Kawada}
\author[3]{T.~Matsuda}
\author[5]{R.D.~Settles}
\author[1]{A.~Sugiyama}
\author[8]{T.~Takahashi}
\author[3]{J.~Tian}
\author[6]{T.~Watanabe}
\author[9]{R.~Yonamine}
%
%
\address[3]{High Energy Accelerator Research Organization (KEK), Tsukuba 305-0801, Japan}
\address[7]{The Graduate University for Advanced Studies (Sokendai), Tsukuba 305-0801, Japan}
\address[8]{Hiroshima University, Higashi-Hiroshima 739-8530, Japan}
\address[1]{Saga University, Saga 840-8502, Japan}
\address[4]{Kinki University, Higashi-Osaka 577-8502, Japan}
\address[10]{Deutsches Elektronen-Synchrotron (DESY), D-22607 Hamburg, Germany}
\address[5]{Max Planck Institute for Physics, DE-80805 Munich, Germany}
\address[6]{Kogakuin University, Shinjuku 163-8677, Japan}
\address[9]{Universit$\acute{e}$ Libre de Bruxelles, 1050 Brussels, Belgium}

\end{frontmatter}

\vskip12pt

\setcounter{figure}{0}


\end{comment}

\begin{center}
\appendix{\underline{{\bf Addendum} to be published in Nucl. Instrum. Methods Phys. Res. A}}
\end{center}

\vspace{2mm}

%
%
%
%

As mentioned in the last paragraph of Appendix A,
the values of the relative variance of gas gain ($f$), shown in Figs.~5--7, are a little smaller than
the real values because of the charge sharing between the neighboring pad rows
due to the spread of multiplied electrons in the GEM stack.

The charge sharing, if it is not negligible, reduces the fluctuations in the number of multiplied electrons collected
by a pad row ($N$), and
Eq.~(A.7) has to be corrected with a factor $g(\sigma/h)$ ($\le 1$):
\begin{equation}
\sigma_{\rm N}^2 \equiv \left< \left( N - \left< N \right> \right)^2 \right> 
= \left< n \right> \cdot \left< g \right>^2 \cdot (1 + f)
  \cdot g\left( \frac{\sigma}{h} \right)
\end{equation}
with
\begin{equation}
  g\left( \frac{\sigma}{h} \right)
  = {\rm erf}\left(\frac{h}{2 \sigma}\right)
      - \frac{2}{\sqrt{\pi}} \cdot \frac{\sigma}{h} \cdot
      \left[ 1 - {\rm exp} \left( - \frac{h^2}{4 \sigma^2}\right)
               \right]
\end{equation}
where $\left< n \right>$ denotes the average number of primary electrons per pad row,
$\left< g \right>$ represents the average gas gain,
$h$ is the pad-row height, and $\sigma$ is the standard deviation of the 
spread (lateral diffusion) of multiplied electrons in the GEM stack.
The function $g(\sigma/h)$ should not be confused with the gas gain ($g$).
On the other hand, $\left< N \right>$ is not affected by the charge sharing.

The basic equation (A.23) then becomes
\begin{equation}
  \left< \left( N_1 - N_2 \right)^2 \right> =
  2 \cdot \left< \left< n \right> \right> \cdot \left< g \right>^2
  \cdot (1+f)
  \cdot g\left( \frac{\sigma}{h} \right)\;.
\end{equation}
Consequently, Eq.~(A.26) and the corresponding equation in the main text (Eq.~(1)) need to be modified:
\begin{equation}
1 + f
= \frac{\left<\left< n \right>\right>}{2} \cdot 
\frac{\left< \left( Q_1 - Q_2 \right)^2
  \right>}{\left<\left< Q \right>\right>^2}
  \cdot \frac{1}{g\left( \frac{\sigma}{h} \right)}
\end{equation}
where $Q_1$ ($Q_2$) represents the charge proportional to $N_1$ ($N_2$) recorded by the readout electronics.

It should be noted that the adjacent pad rows are located in the middle of the $\sim$200-mm-long waist of the laser beam,
where photons travel in parallel with the beam axis.
Hence, we assume $\left< n \right>$ defined by Eq.~(A.8) for each of the laser shots is the same for the two pad rows.
The observed small difference in
$\left<\left< Q \right>\right> \propto \left<\left< N \right>\right> = \left< g \right> \cdot \left<\left< n \right>\right>$
is, therefore, attributed to the different average gas gains $\left<g_1\right>$ and $\left<g_2\right>$.
The signal charge $Q_2$ for each laser shot is scaled by a constant factor beforehand so that  
$\left<\left< Q \right>\right>$ $\equiv$ $\left<\left< Q_1 \right>\right>$ = $\left<\left< Q_2 \right>\right>$.
It is also to be noted that Fig.~2 is misleading because we actually used a pair of alternating pad rows
instead of neighboring ones (see text).

When $\sigma \ll h$, as is the case for Micromegas, for example,  
$g(\sigma / h)$ is close to unity,
and the correction for the charge sharing is negligible.

See Section~6.1 and Appendix~B of Ref.~\cite{Kobayashi} for details and the corrected values of $f$.


\end{document}